\documentclass[11pt]{article}
\usepackage[dvips]{graphicx}  
\begin{document}

\title{\LARGE \bf Quantum-Foam  In-Flow  Theory of Gravity and the Global Positioning 
System (GPS) }  
\author{Reginald T. Cahill\footnote{{\bf Process Physics:}\newline
        http://www.mountainman.com.au/process$\_$physics/ \newline
http://www.scieng.flinders.edu.au/cpes/people/cahill\_r/processphysics.html }\\
 {School of Chemistry, Physics and Earth Sciences}\\
{ Flinders University }\\ 
{ GPO Box 2100, Adelaide 5001, Australia }\\
{(Reg.Cahill@flinders.edu.au)}}

\date{}
\maketitle

\begin{abstract}
It is shown that a new quantum-foam in-flow theory of gravity is mathematically equivalent to the General Relativity theory of
gravity for the operation of the Global Positioning System (GPS). The differences between the two theories 
become experimentally evident  in other situations such as in the so-called `dark matter' effect, in the 
observation of absolute motion and {\it ipso facto} in the observation of the in-flow motion into the sun, and in the observation of
a new class of  gravitational waves, effects  which are present in   existing experimental observations, but are {\it not} within
General Relativity.  This new theory of gravity arises within the information-theoretic {\it Process Physics}.
\end{abstract}

\begin{center}
\vspace{20mm}

{\bf - September 2003 -  Updated July 2004 -}

\vspace{5mm}

 { \bf \large arXiv:physics/0309016}
\end{center}

\newpage

\tableofcontents

\newpage

\section{ Introduction\label{section:introduction}}

It has been  extensively argued that the very successful operation of the Global Positioning System (GPS)
\cite{Ashby} is proof of the validity of the General Relativity formalism for gravity.  However as is well known,
and was most clearly stated by Popper, in science agreement with observation does not amount to the proof of the
theory used to successfully describe the experimental  data; in fact experiment can only strictly be used to
disprove a theory. We illustrate this herein by discussing a very different theory of gravity in which
gravitational forces are caused by inhomogeneities in the effective in-flow of the quantum-foam substratum, that
is space, into matter.   We shall show that this new theory of gravity and General Relativity are  mathematically
equivalent when it comes to explaining the operation of the GPS, because of special circumstances prevailing in
this case\footnote{The key parts of this paper are
considerably changed from the earlier version of September 2003. Since that time the new theory of gravity has
developed considerably, and the analysis regarding the  approximate velocity `flow' superposition principle, and
the mapping of the new theory of gravity, in the case of the circumstances for the GPS, back to the corresponding
General Relativity mathematical formalism treatment, is considerably enhanced.}.    The predictive differences
between the two theories  become experimentally evident  in other situations such as in the so-called `dark matter'
effect, in the  observation of absolute motion and {\it ipso facto} in the observation of the in-flow  into the
sun, and in the observation of a new class of  gravitational waves, and various other known `gravitational
anomalies', effects  which are present in   existing experimental observations, but are {\it not} in General
Relativity.  

The new theory of gravity arises within the information-theoretic {\it Process Physics}
\cite{NovaBook,RC01,CK98,CK99,CKK00,CK,GQF,AMGE,RGC,alpha,NovaDM,GPB}. 
Here we proceed  by showing that both the  Newtonian theory of gravity and General Relativity, in the
case of the external Schwarzschild metric relevant to the GPS, may be written as in-flow dynamical systems.  A
generalisation of these in-flow formalisms is proposed giving the   new theory of gravity.  This theory
possesses distinct and observable effects already evident in existing experimental data. However in the
case of high spherical symmetry, which is relevant to the GPS, the new theory of gravity becomes
mathematically equivalent to General Relativity with an external Schwarzschild metric, but has  a vastly different
interpretation and ontology.   The key insight is that the dynamical effects of the detectable  motion
through the quantum-foam substratum causes relativistic effects and these, together with the
quantum-foam in-flow effects, explain the operation of the GPS. 

As discussed in \cite{AMGE} numerous interferometer and non-interferometer experiments have detected absolute
motion of the Solar system   in the direction (Right Ascension $=5.2^{hr}$, Declination$ =
-67^0$)  with a speed of
$417 \pm 40$ km/s. This is the velocity after  removing the contribution of the earth's
orbital speed and the sun in-flow effect. It is significant that this velocity is different
to that associated with the Cosmic Microwave Background  (CMB) relative to which the Solar system
has a speed of $369$ km/s in the direction $(\alpha=11.20^h,\delta=-7.22^0)$, see \cite{CMB}. As well the
experimental data also reveals an in-flow of space past the earth towards the sun.  It needs to be emphasised
that the detection of absolute motion is fully consistent with the well known special relativity effects, and
indeed these effects, namely time dilations and length contractions, are needed to understand the operation in
particular of the Michelson interferometer, which formed the basis of several key experiments.  The   major
insight is that absolute motion through the quantum foam substratum is the cause of the special relativistic
effects.  The detection of this absolute motion is evidence that space has structure though the scale of that
structure is not revealed by the experiments analysed in \cite{AMGE}.

\section[       Gravity  as Inhomogeneous Quantum Foam In-Flow ]{ Gravity as Inhomogeneous Quantum Foam In-Flow 
\label{section:QFinflow}} 
Here we show that the Newtonian theory of gravity may be exactly re-written as a `fluid flow' system,
as can General Relativity for a class of metrics.  This `fluid' system is interpreted \cite{NovaBook,
RC01} as a classical description of a quantum foam substructure to space, and the `flow' describes the
relative motion of this quantum foam with, as we now show, gravity arising from inhomogeneities in that
flow. These inhomogeneities can be caused by an in-flow into matter, or even as inhomogeneities
produced purely by the self-interaction of space itself, as happens for instance  for the black holes.  

The Newtonian
theory was originally formulated in terms of a force field, the gravitational acceleration
${\bf g}({\bf r},t)$, which is determined by  the matter density
$\rho({\bf r},t)$ according to
\begin{equation}\label{eqn:g1}
\nabla.{\bf g}=-4\pi G\rho.
\end{equation}
For $\nabla \times {\bf g}=0$ this gravitational acceleration ${\bf g}$ may be written as the
gradient of the gravitational potential $\Phi$ 
\begin{equation}{\bf g}=-{\bf \nabla}\Phi,\label{eqn:gPhi}\end{equation}  
where the  gravitational
potential\index{gravitational potential} is now determined  by 
\begin{equation}
 \nabla^2\Phi=4\pi G\rho .
\end{equation} 
 Here, as usual, $G$ is
the Newtonian  gravitational constant.  Now as $\rho\geq 0$ we can choose to have 
$\Phi
\leq 0$ everywhere if $\Phi \rightarrow 0$ at infinity. So we can introduce  ${\bf v}^2=-2\Phi \geq 0$ where 
$\bf v$ is some velocity vector field.
  Here the value of ${\bf v}^2$ is
specified, but not the direction of ${\bf v}$. Then
\begin{equation}
{\bf g}=\frac{1}{2}{\bf \nabla}({\bf v}^2)=({\bf v}.{\bf \nabla}){\bf v}+
{\bf v}\times({\bf \nabla}\times{\bf v}).
\label{eqn:f1}
\end{equation} 
For zero-vorticity (irrotational) flow 
${\bf \omega}={\bf \nabla} \times {\bf v}={\bf 0}$. Then ${\bf g}$ is
the usual Euler expression for the  acceleration of a fluid element in a
time-independent or stationary fluid flow.   If the flow is time dependent that expression
is expected to become
 \begin{equation}{\bf g}=\displaystyle{\frac{\partial {\bf v}}{\partial t}}+({\bf v}.{\bf \nabla}){\bf
v}=\displaystyle{\frac{d {\bf v}}{d t}},
\label{eqn:f2}\end{equation}
which has given rise to the total derivative of ${\bf v}$ familiar from fluid mechanics. This equation  is then to
be accompanied by the `Newtonian equation' for the flow field
\begin{equation}
\frac{1}{2}\nabla^2({\bf v}^2)=-4\pi G\rho,
\label{eqn:f3a}\end{equation}
but to be consistent with (\ref{eqn:f2}) in the case of a time-dependent matter density this equation
should be generalised to 
\begin{equation}
\frac{\partial }{\partial t}(\nabla.{\bf v})+\nabla.(({\bf
v}.{\bf \nabla}){\bf v})=-4\pi G\rho.
\label{eqn:f3}\end{equation}
This exhibits the fluid flow form of  Newtonian gravity in the case of zero vorticity $\nabla \times
{\bf v}=0$.  For zero vorticity (\ref{eqn:f3}) determines both the magnitude and direction of the
velocity field, for in this case we can write ${\bf v}=\nabla u$, where $u({\bf r},t)$ is a scalar
velocity potential, and in terms of $u({\bf r},t)$ (\ref{eqn:f3})  specifies uniquely the time evolution
of $u({\bf r},t)$.  Note that (\ref{eqn:f2})  and (\ref{eqn:f3})  are exactly equivalent to
(\ref{eqn:g1}) for the acceleration field ${\bf g}$, and so   within the fluid flow formalism
(\ref{eqn:f2}) and (\ref{eqn:f3}) are together equivalent to the Universal Inverse Square Law for
 ${\bf g}$, and so both are equally valid as regards the numerous experimental and observational checks
of the acceleration field ${\bf g}$  formalism, particularly the Keplerian rotation velocity law. So we
appear to have two equivalent formalisms for the same phenomenon. Indeed for a stationary spherically symmetric
distribution of matter of total mass
$M$ the velocity field outside of the matter 
\begin{equation}
{\bf v}({\bf r})=-\sqrt{\frac{2GM}{r}}\hat{\bf r},
\label{eqn:vfield}\end{equation}
satisfies (\ref{eqn:f3}) and reproduces the inverse square law form for ${\bf g}$ using (\ref{eqn:f2}): 
\begin{equation}
{\bf g}=-\frac{GM}{r^2}\hat{\bf r}.
\label{eqn:InverseSqLaw}\end{equation} 

So the immediate questions that arise are (i)  can the two formalisms be distinguished experimentally, and (ii)
can the velocity field formalism be generalised, leading  to new gravitational phenomena?   To answer these questions we
note that 
\begin{enumerate}
\item The velocity flow field of some $417\pm40$ km/s in the direction (Right Ascension $=5.2^{hr}$, Declination$ =
-67^0$)  has been detected in several experiments, as described in considerable detail in 
\cite{NovaBook,AMGE,RGC}. The major component of that flow is related to a galactic flow, presumably  within the
Milky Way and  the local galactic cluster, but a smaller component of some 50km/s being the  flow past the earth
towards the sun has also recently been revealed in the data.

\item In terms of the velocity field formalism (\ref{eqn:f3}) a unique term may be added that does not affect
observations  within the solar system, such as encoded in Kepler's laws, but outside of that special case the new term
causes effects which vary from small to extremely large.  This term will be shown herein to cause those effects that have
been mistakenly called the `dark matter' effect. 

\item Eqn.(\ref{eqn:f3}) and its generalisations have time-dependent solutions even when the matter density is not
time-dependent. These are a form of flow turbulence, a gravitational wave effect, and they have also been detected, as
discussed in \cite{NovaBook,AMGE,RGC}. 

\item  The need for a further generalisation of the flow equations will be argued for, and this in particular
includes flow vorticity that leads to a non-spacetime explanation of the `frame-dragging' effect, and of the `dark
matter' network observed using the weak gravitational lensing technique.
\end{enumerate}

First let us consider the arguments that lead to a generalisation of (\ref{eqn:f3}). The simplest generalisation
is
\begin{equation}
\frac{\partial }{\partial t}(\nabla.{\bf v})+\nabla.(({\bf
v}.{\bf \nabla}){\bf v})+C({\bf v})=-4\pi G\rho,
\label{eqn:f3extend}\end{equation}
where
\begin{equation}
C({\bf v})=\displaystyle{\frac{\alpha}{8}}((tr D)^2-tr(D^2)),
\label{eqn:Cdefn1}\end{equation} and
\begin{equation}
D_{ij}=\frac{1}{2}(\frac{\partial v_i}{\partial x_j}+\frac{\partial v_j}{\partial x_i})
\label{eqn:Ddefn1}\end{equation}
is the symmetric part of the rate of strain tensor $\partial v_i/\partial x_j$, and $\alpha$ is a dimensionless
constant - a new gravitational constant in addition to $G$.  It is possible to check that for the in-flow in
(\ref{eqn:vfield}) $C({\bf v})=0$. This is a feature that uniquely determines the form of $C({\bf v})$.  This means
that effects  caused by this new term are not manifest in the planetary motions that formed the basis of Kepler's
phenomenological laws and that then lead to Newton's theory of gravity.  As discussed in \cite{alpha,NovaDM} the
value of $\alpha$ determined from the Greenland bore hole $g$ anomaly experimental data is found to be the fine
structure constant, to within experimental error. As  well, as discussed in \cite{alpha,NovaDM}
(\ref{eqn:f3extend}) predicts precisely the  so-called `dark matter' effect, with the effective `dark matter'
density defined by  
\begin{equation}
\rho_{DM}({\bf r})=\frac{\alpha}{32\pi G}( (tr D)^2-tr(D^2)).  
\label{eqn:DMdensity0}\end{equation} 
So the explanation of the `dark matter'
effect becomes apparent  once we use the velocity field  formulation of gravity. However  (\ref{eqn:f3extend}) must
be further generalised to include  (i)  the velocity of absolute motion  of the matter
components with respect to the local quantum foam system, and (ii) vorticity effects.

For these further generalisations\footnote{The remainder of this section has been considerably changed since the
September 2003 version of this paper.} we need to be  precise by what is meant by the velocity field
 ${\bf v}({\bf r},t)$. To be specific and also to define a possible measurement procedure
 we can choose to  use the Cosmic Microwave Background (CMB)\index{Cosmic Microwave
Background (CMB)} frame of reference for that purpose, as this is itself easy to establish. However that does not
imply that the  CMB frame is the local `quantum-foam' rest frame. Relative to the CMB frame and using the local
absolute motion detection techniques described in \cite{NovaBook,AMGE,RGC}, or more modern techniques that are
under development, ${\bf v}({\bf r},t)$ may be measured in the neighbourhood of the observer.   Then an `object' at
location ${\bf r}_0(t)$ in the CMB frame has velocity  
 ${\bf v}_0(t)=d{\bf r}_0(t)/dt$ 
with respect to that  frame.  We then define 
\begin{equation}
{\bf v}_R({\bf r}_0(t),t) ={\bf v}_0(t) - {\bf v}({\bf r}_0(t),t),
\label{eqn:18}
\end{equation}
as the velocity of the object relative to the quantum foam at the location of the object.  However this absolute
velocity of matter ${\bf v}_R(t)$ does not appear in (\ref{eqn:f3extend}), and so not only is that equation lacking
vorticity effects, it presumably is only an approximation for when the matter has a negligible speed of absolute
motion with respect to the local quantum foam.  To introduce  the vector ${\bf v}_R(t)$ we need to construct a
2nd-rank tensor generalisation of   (\ref{eqn:f3extend}), and the simplest form is 
\begin{eqnarray}
&&\frac{d D_{ij}}{dt}+\frac{\delta_{ij}}{3}tr(D^2) +\frac{tr D}{2}
(D_{ij}-\frac{\delta_{ij}}{3}tr D)\nonumber \\ &&+\frac{\delta_{ij}}{3}\frac{\alpha}{8}((tr D)^2 -tr(D^2))=-4\pi
G\rho(\frac{\delta_{ij}}{3}+\frac{v^i_{R}v^j_{R}}{2c^2}+..)  ,  \mbox{ } i,j=1,2,3.
\label{eqn:f3general}\end{eqnarray}
which uses the total derivative of the $D_{ij}$ tensor in (\ref{eqn:Ddefn1}). Because of its tensor structure we
can  now include the direction of absolute motion of the matter density with respect to the quantum foam, with the
scale of that given by
$c$, which is the speed of light relative to the quantum foam. The superscript notation for the components of   ${\bf
v}_R(t)$ is for convenience only, and has no other significance.  The trace of (\ref{eqn:f3general}), using the
identity
\begin{equation}
({\bf v}.\nabla)(tr D)=\frac{1}{2}\nabla^2({\bf v}^2)
-tr(D^2)-\frac{1}{2}(\nabla\times{\bf v})^2+ {\bf v}.\nabla\times(\nabla\times{\bf v}),
\label{eqn:identity}\end{equation}
gives, for zero vorticity,
\begin{equation}
\frac{\partial }{\partial t}(\nabla.{\bf v})+\nabla.(({\bf
v}.{\bf \nabla}){\bf v})+C({\bf v})=-4\pi G\rho(1+\frac{v_R^2}{2c^2}+..),
\label{eqn:f3p}\end{equation}
which is (\ref{eqn:f3extend}) in the limit $v_R\rightarrow 0$.  As well the off-diagonal terms,  $i\neq j$,
are satisfied, to   $O(v_R^iv_R^j/c^2)$, for the in-flow velocity field in (\ref{eqn:vfield}). The
conjectured form of the RHS of (\ref{eqn:f3p}) is, to 
$O(v_R^2/c^2)$,  based on the Lorentz contraction effect for the matter density, with
$\rho$ defined as the matter density  if the matter were at rest with respect to the quantum foam. Hence, because of
(\ref{eqn:f3p}),  (\ref{eqn:f3general})  is in agreement with Keplerian orbits for the solar system with the
velocity field given by   (\ref{eqn:vfield}).

We now consider a further generalisation of (\ref{eqn:f3general}) to include vorticity effects, namely 
\begin{eqnarray}
&&\frac{d D_{ij}}{dt}+ \frac{\delta_{ij}}{3}tr(D^2) + \frac{tr D}{2}
(D_{ij}-\frac{\delta_{ij}}{3}tr D)\nonumber \\ &&+\frac{\delta_{ij}}{3}\frac{\alpha}{8}((tr
D)^2 -tr(D^2)) -(D\Omega-\Omega D)_{ij}\nonumber \\&&\mbox{\ \ \ \ \ \ \ \ \  }=-4\pi
G\rho(\frac{\delta_{ij}}{3}+\frac{v^i_{R}v^j_{R}}{2c^2}+..), 
\mbox{ \ \ } i,j=1,2,3, 
\label{eqn:f3vorticitya}\end{eqnarray}
\begin{equation}\nabla \times(\nabla\times {\bf v}) =\frac{8\pi G\rho}{c^2}{\bf v}_R,
\label{eqn:f3vorticityb}\end{equation}
 where
\begin{equation}
\Omega_{ij}=\frac{1}{2}(\frac{\partial v_i}{\partial x_j}-\frac{\partial v_j}{\partial
x_i})=-\frac{1}{2}\epsilon_{ijk}\omega_k=-\frac{1}{2}\epsilon_{ijk}(\nabla\times {\bf v})_k
\label{eqn:Omegadefn}\end{equation}
is the antisymmetric part of the rate of strain tensor $\partial v_i/\partial x_j$, which is the vorticity vector
field $\omega$ in tensor form. The term
$(D\Omega-\Omega D)_{ij}$ allows the vorticity  vector field to be coupled to the symmetric tensor $D_{ij}$
dynamics. Again the vorticity is generated by absolute motion of the matter density with respect to the local
quantum foam. Eqns (\ref{eqn:f3vorticitya}) and (\ref{eqn:f3vorticityb}) now permit the time evolution of the
velocity field  to be determined. Note that the vorticity equation in (\ref{eqn:f3vorticityb}) may be explicitly
solved, for it may be written as
\begin{equation}
\nabla(\nabla.{\bf v})-\nabla^2 {\bf v}=\frac{8\pi G\rho}{c^2}{\bf v}_R,
\label{veqn}\end{equation}
which gives, using
\begin{equation}
\nabla^2\left(\frac{1}{|{\bf r} - {\bf r^\prime}|}  \right)=-4\pi\delta({\bf r} - {\bf r^\prime}),
\label{eqn:deltafnidentity}\end{equation}
\begin{equation}
{\bf v}({\bf r},t)=\frac{2G}{c^2}\int d^3 r^\prime \frac{\rho({\bf r}^\prime,t)}{|{\bf r}-{\bf r}^\prime|}{\bf
v}_R({\bf r}^\prime,t)-
\frac{1}{4\pi}\int d^3 r^\prime \frac{1}{|{\bf r}-{\bf r}^\prime|}\nabla(\nabla.{\bf v}({\bf
r}^\prime,t)).
\label{eqn:BS1}\end{equation}
This suggests that ${\bf v}({\bf r},t)$ is now determined solely by the vorticity equation. However  (\ref{eqn:BS1})
is misleading, as  (\ref{eqn:f3vorticityb}) only specifies the vorticity, and taking the $\nabla \times$ of  
(\ref{eqn:BS1}) we obtain
\begin{equation}
\omega({\bf r},t)=\nabla\times{\bf v}({\bf r},t)
=\frac{2G}{c^2}\int d^3 r^\prime \frac{\rho({\bf r}^\prime,t)}
{|{\bf r}-{\bf r}^\prime|^3}{\bf v}_R({\bf r}^\prime,t)\times({\bf r}-{\bf r}^\prime)+\nabla\psi,
\label{eqn:BS2}\end{equation} 
which is the Biot-Savart form for the vorticity, with the additional term being the homogeneous solution. The
homogeneous term corresponds to (distant) matter densities not explicitly included in $\rho({\bf r}^\prime,t)$.
Then (\ref{eqn:f3vorticitya}) becomes an integro-differential equation for the velocity field.   As discussed  in
\cite{GPB}   (\ref{eqn:BS2}) explains the so-called 
`frame-dragging'  effect in terms of this vorticity in the in-flow, but makes predictions very different from
General Relativity.  These conflicting predictions will soon be tested by the Gravity Probe B satellite
experiment.   Of course (\ref{eqn:f3vorticitya}) and (\ref{eqn:BS2}) only make sense if  
${\bf v}_R({\bf r},t)$ for the matter at location ${\bf r}$  is specified. We now consider the special case where
the matter is subject only to the effects of motion with respect to the quantum-foam velocity-field
inhomogeneities and  variations in time, which causes a `gravitational' acceleration.

We note that the first serious attempt to construct a `flow' theory of gravity was by Kirkwood \cite{RK1,RK2}.
However the above theory, as expressed in (\ref{eqn:f3vorticitya}) and (\ref{eqn:f3vorticityb}), is very different
to Kirkwood's proposal. We also note that (\ref{eqn:f3vorticitya}) and (\ref{eqn:f3vorticityb}) need to be further
generalised  to take account of the cosmological-scale effects, namely that the spatial system is compact and
growing, as discussed in
\cite{NovaBook}.

\section{ Geodesics \label{section:geodesics}}

Process Physics \cite{NovaBook} leads to the Lorentzian interpretation\index{Lorentzian interpretation} of
so called `relativistic effects'.  This means that the speed of light \index{speed of light} is only `c' with
respect to the quantum-foam system, and that time dilation effects for clocks and length contraction effects for
rods are caused by the motion of clocks and rods relative to the quantum foam. So these effects are real dynamical
effects caused by motion through the quantum foam, and are not to be interpreted as non-dynamical spacetime
effects as suggested by Einstein.  To arrive at the dynamical description of the various effects of the quantum foam
we shall introduce conjectures that essentially lead to a phenomenological description of these effects. In
the future we expect to be able to derive this dynamics directly from the Quantum Homotopic Field Theory (QHFT)
 that describes the quantum foam system \cite{NovaBook}. Here we shall conjecture that the path of an object
through an inhomogeneous and time-varying quantum-foam is determined, at a classical level,  by a variational
principle, namely that  the travel time is extremised for the physical path ${\bf r}_0(t)$. The travel time is
defined by 
\begin{equation}
\tau[{\bf r}_0]=\int dt \left(1-\frac{{\bf v}_R^2}{c^2}\right)^{1/2},
\label{eqn:f4}
\end{equation}  
with ${\bf v}_R$ given by (\ref{eqn:18}). So the trajectory will be independent of the mass of the object,
corresponding to the equivalence principle.  Under a deformation of the trajectory  ${\bf r}_0(t) \rightarrow  {\bf
r}_0(t) +\delta{\bf r}_0(t)$,
${\bf v}_0(t) \rightarrow  {\bf v}_0(t) +\displaystyle\frac{d\delta{\bf r}_0(t)}{dt}$,  and we also
have
\begin{equation}\label{eqn:G2}
{\bf v}({\bf r}_0(t)+\delta{\bf r}_0(t),t) ={\bf v}({\bf r}_0(t),t)+(\delta{\bf
r}_0(t).{\bf \nabla}) {\bf v}({\bf r}_0(t))+... 
\end{equation}
Then
\begin{eqnarray}\label{eqn:G3}
\delta\tau&=&\tau[{\bf r}_0+\delta{\bf r}_0]-\tau[{\bf r}_0]  \nonumber\\
&=&-\int dt \:\frac{1}{c^2}{\bf v}_R. \delta{\bf v}_R\left(1-\displaystyle{\frac{{\bf
v}_R^2}{c^2}}\right)^{-1/2}+...\nonumber\\
&=&\int dt\frac{1}{c^2}\left({\bf
v}_R.(\delta{\bf r}_0.{\bf \nabla}){\bf v}-{\bf v}_R.\frac{d(\delta{\bf
r}_0)}{dt}\right)\left(1-\displaystyle{\frac{{\bf v}_R^2}{c^2}}\right)^{-1/2}+...\nonumber\\ 
&=&\int dt \frac{1}{c^2}\left(\frac{{\bf v}_R.(\delta{\bf r}_0.{\bf \nabla}){\bf v}}{ 
\sqrt{1-\displaystyle{\frac{{\bf
v}_R^2}{c^2}}}}  +\delta{\bf r}_0.\frac{d}{dt} 
\frac{{\bf v}_R}{\sqrt{1-\displaystyle{\frac{{\bf
v}_R^2}{c^2}}}}\right)+...\nonumber\\
&=&\int dt\: \frac{1}{c^2}\delta{\bf r}_0\:.\left(\frac{({\bf v}_R.{\bf \nabla}){\bf v}+{\bf v}_R\times({\bf
\nabla}\times{\bf v})}{ 
\sqrt{1-\displaystyle{\frac{{\bf
v}_R^2}{c^2}}}}  +\frac{d}{dt} 
\frac{{\bf v}_R}{\sqrt{1-\displaystyle{\frac{{\bf
v}_R^2}{c^2}}}}\right)+...
\end{eqnarray}
  Hence a 
trajectory ${\bf r}_0(t)$ determined by $\delta \tau=0$ to $O(\delta{\bf r}_0(t)^2)$ satisfies 
\begin{equation}\label{eqn:G4}
\frac{d}{dt} 
\frac{{\bf v}_R}{\sqrt{1-\displaystyle{\frac{{\bf v}_R^2}{c^2}}}}=-\frac{({\bf
v}_R.{\bf \nabla}){\bf v}+{\bf v}_R\times({\bf
\nabla}\times{\bf v})}{ 
\sqrt{1-\displaystyle{\frac{{\bf v}_R^2}{c^2}}}}.
\label{eqn:vReqn}\end{equation}
Let us now write this in a more explicit form.  This will
also allow the low speed limit to be identified.   Substituting ${\bf
v}_R(t)={\bf v}_0(t)-{\bf v}({\bf r}_0(t),t)$ and using 
\begin{equation}\label{eqn:G5}
\frac{d{\bf v}({\bf r}_0(t),t)}{dt}=\frac{\partial {\bf v}}{\partial t}+({\bf v}_0.{\bf \nabla}){\bf
v},
\end{equation}
we obtain
\begin{equation}\label{eqn:G6}
\frac{d}{dt} 
\frac{{\bf v}_0}{\sqrt{1-\displaystyle{\frac{{\bf v}_R^2}{c^2}}}}={\bf v}
\frac{d}{dt}\frac{1}{\sqrt{1-\displaystyle{\frac{{\bf v}_R^2}{c^2}}}}+\frac{\displaystyle{\frac{\partial {\bf
v}}{\partial t}}+({\bf v}.{\bf \nabla}){\bf v}+({\bf \nabla}\times{\bf v})\times{\bf v}_R}{ 
\displaystyle{\sqrt{1-\frac{{\bf v}_R^2}{c^2}}}}.
\end{equation}
Then in the low speed limit  $v_R \ll c $   we  obtain
\begin{equation}{\label{eqn:G7}}
\frac{d{\bf v}_0}{dt}=\frac{\partial {\bf v}}{\partial t}+({\bf v}.{\bf
\nabla}){\bf v}+({\bf \nabla}\times{\bf v})\times{\bf v}_R,
\end{equation}
which agrees with the  fluid flow form suggested in  (\ref{eqn:f2}) for zero vorticity (${\bf \nabla}\times{\bf
v}=0$), but introduces a new vorticity effect for the gravitational acceleration.  The last term in (\ref{eqn:G7}) is
relevant to the `frame-dragging' effect and  to the Allais eclipse effect. Hence (\ref{eqn:G6}) is a
generalisation of (\ref{eqn:f2}) to include  Lorentzian dynamical effects, for  in (\ref{eqn:G6})  we can
multiply both sides by the rest mass  $m_0$ of the object, and   then (\ref{eqn:G6}) involves 
\begin{equation}
m({\bf v}_R) =\frac{m_0}{\sqrt{1-\displaystyle{\frac{{\bf v}_R^2}{c^2}}}},
\label{eqn:G8}\end{equation}
the so called `relativistic' mass\index{relativistic mass}, and (\ref{eqn:G6}) acquires the form
\begin{equation}\frac{d}{dt}(m({\bf v}_R){\bf v}_0)={\bf F},\end{equation} where
${\bf F}$ is an effective `force' caused by the inhomogeneities and time-variation of the flow.  This is
essentially Newton's 2nd Law of Motion in the case of gravity only. That $m_0$ cancels is the equivalence principle, 
and which acquires a simple explanation in terms of the flow.  Note that the
occurrence of
$1/\sqrt{1-\frac{{\bf v}_R^2}{c^2}}$ will lead to the precession of the perihelion of elliptical planetary orbits,
and also to horizon effects wherever  $|{\bf v}| = c$: the region where  $|{\bf v}| < c$ is
inaccessible from the region where $|{\bf v}|>c$.  Also (\ref{eqn:f4}) is easily used to determine the 
clock rate offsets in the GPS satellites\index{Global Positioning System (GPS)}, when the in-flow is given by
(\ref{eqn:vfield}). So the fluid flow dynamics in  (\ref{eqn:f3vorticitya}) and (\ref{eqn:BS2}) and the gravitational
dynamics for the matter in  (\ref{eqn:vReqn}) now form a closed system.  
This system of equations is a considerable
generalisation from that of Newtonian gravity, and would appear to be very different from the curved spacetime
formalism of General Relativity. However we now show that General Relativity leads to a very similar system of
equations, but with one important exception, namely that the `dark matter' `quantum-foam' dynamics is missing from the
Hilbert-Einstein theory of gravity. 

The above may  be modified when the `object' is a massless photon, and the
corresponding result leads to the gravitational lensing effect. But not only will ordinary matter produce such
lensing, but the effective `dark matter' density will also do so, and that is relevant to the recent observation by
the weak lensing technique of the so-called `dark matter' networks.

\section[  General Relativity and the In-Flow Process ]{  General Relativity and the In-Flow Process 
\label{section:general}} 

Eqn.(\ref{eqn:f4})  involves various absolute quantities such  as the absolute velocity of an object
relative to the  quantum foam and the absolute speed
$c$ also relative to the foam, and of course absolute velocities are excluded from the General Relativity (GR)
formalism.  Here we examine GR to point out the key differences with the new theory, but also to indicate why, in
the special case of the external-Scharwzschild metric, GR was apparently but misleadingly succesful.  In particular
we find that the major failing of GR is that it was constructed to agree with the Newtonian theory in the limit of
low velocities, and so  by default excluded the `dark matter' effect. However (\ref{eqn:f4}) gives (with
$t=x_0^0$)
\begin{equation}
d\tau^2=dt^2-\frac{1}{c^2}(d{\bf r}_0(t)-{\bf v}({\bf r}_0(t),t)dt)^2=
g_{\mu\nu}(x_0 )dx^\mu_0 dx^\nu_0,
\label{eqn:24}\end{equation}
which is  the   Panlev\'{e}-Gullstrand form of the metric $g_{\mu\nu}$
\index{metric $g_{\mu\nu}$} 
\cite{PP, AG} for GR.  We emphasize that the
absolute velocity
${\bf v}_R$  has been measured, and so the foundations of GR as usually stated are invalid. 
Here we look closely at the GR formalism when the metric has the form in (\ref{eqn:24}), appropriate to a 
velocity field  formulation of gravity.  In GR the  metric tensor $g_{\mu\nu}(x)$, specifying the geometry
of the spacetime construct, is determined by
\begin{equation}
G_{\mu\nu}\equiv R_{\mu\nu}-\frac{1}{2}Rg_{\mu\nu}=\frac{8\pi G}{c^2} T_{\mu\nu},
\label{eqn:32}\end{equation}
where  $G_{\mu\nu}$ is  the Einstein tensor, $T_{\mu\nu}$ is the  energy-momentum tensor,
$R_{\mu\nu}=R^\alpha_{\mu\alpha\nu}$ and
$R=g^{\mu\nu}R_{\mu\nu}$ and
$g^{\mu\nu}$ is the matrix inverse of $g_{\mu\nu}$. The curvature tensor is
\begin{equation}
R^\rho_{\mu\sigma\nu}=\Gamma^\rho_{\mu\nu,\sigma}-\Gamma^\rho_{\mu\sigma,\nu}+
\Gamma^\rho_{\alpha\sigma}\Gamma^\alpha_{\mu\nu}-\Gamma^\rho_{\alpha\nu}\Gamma^\alpha_{\mu\sigma},
\label{eqn:curvature}\end{equation}
where $\Gamma^\alpha_{\mu\sigma}$ is the affine connection
\begin{equation}
\Gamma^\alpha_{\mu\sigma}=\frac{1}{2} g^{\alpha\nu}\left(\frac{\partial g_{\nu\mu}}{\partial x^\sigma}+
\frac{\partial g_{\nu\sigma}}{\partial x^\mu}-\frac{\partial g_{\mu\sigma}}{\partial x^\nu} \right).
\label{eqn:affine}\end{equation}
In this formalism the trajectories of test objects are determined by
\begin{equation}
\Gamma^\lambda_{\mu\nu}\frac{dx^\mu}{d\tau}\frac{dx^\nu}{d\tau}+\frac{d^2x^\lambda}{d\tau^2}=0,
\label{eqn:33}\end{equation}
 which is equivalent to extremising the functional
\begin{equation}
\tau[x]=\int dt\sqrt{g^{\mu\nu}\frac{dx^{\mu}}{dt}\frac{dx^{\nu}}{dt}},
\label{eqn:path}\end{equation}
with respect  to the path $x[t]$. This is precisely equivalent to (\ref{eqn:f4}).  

In the case   of a spherically symmetric mass $M$ the well known   solution of
(\ref{eqn:32}) outside of that mass    is the external-Schwarzschild metric
\begin{equation}
d\tau^2=(1-\frac{2GM}{c^2r})dt^{ 2}-
\frac{1}{c^2}r^{ 2}(d\theta^2+\sin^2(\theta)d\phi^2)-\frac{dr^{ 2}}{c^2(1-\frac{\displaystyle
2GM}{\displaystyle c^2r})}.
\label{eqn:SM}\end{equation}
This solution is the basis of various experimental checks of General Relativity in which the spherically symmetric
mass is either the sun or the earth.  The four tests are: the gravitational redshift, the bending of light, the
precession of the perihelion of Mercury, and the time delay of radar signals.

 However the solution (\ref{eqn:SM}) is in fact
completely equivalent to the in-flow interpretation of Newtonian gravity.  Making the change of variables
$t\rightarrow t^\prime$ and
$\bf{r}\rightarrow {\bf r}^\prime= {\bf r}$ with
\begin{equation}
t^\prime=t+
\frac{2}{c}\sqrt{\frac{2GMr}{c^2}}-\frac{4GM}{c^2}\mbox{tanh}^{-1}\sqrt{\frac{2GM}{c^2r}},
\label{eqn:37}\end{equation}
the Schwarzschild solution (\ref{eqn:SM}) takes the form
\begin{equation}
d\tau^2=dt^{\prime 2}-\frac{1}{c^2}(dr^\prime+\sqrt{\frac{2GM}{r^\prime}}dt^\prime)^2-\frac{1}{c^2}r^{\prime
2}(d\theta^{\prime 2}+\sin^2(\theta^\prime)d\phi^{\prime}),
\label{eqn:PG}\end{equation}
which is exactly  the  Panlev\'{e}-Gullstrand form of the metric $g_{\mu\nu}$ 
\cite{PP, AG} in (\ref{eqn:24})
 with the velocity field given exactly  by the Newtonian form in (\ref{eqn:vfield}).   In which case the geodesic
equation (\ref{eqn:33}) of test objects in the Schwarzschild metric is equivalent to solving (\ref{eqn:G6}).  
This choice of coordinates corresponds to a particular frame of reference in which the test object has velocity
${\bf v}_R={\bf v}-{\bf v}_0$ relative to the in-flow field ${\bf v}$, as seen in (\ref{eqn:f4}).    This results
shows that the Schwarzschild metric in GR is completely equivalent to Newton's inverse square law: GR in
this case is nothing more than Newtonian gravity in disguise.  So the so-called `tests' of GR were nothing more than
a test of the geodesic equation, where most simply this is seen to determine the motion of an object relative to an
absolute local frame of reference - the quantum foam frame. 

 It is conventional wisdom for practitioners in  General
Relativity  to regard the choice of coordinates or frame of reference to be entirely arbitrary and having no physical
significance:  no observations should be possible that can detect and measure ${\bf v}_R$.  This `wisdom' is based
on two  beliefs (i) that all attempts to detect ${\bf v}_R$, namely the detection of absolute motion, have
failed, and that (ii)  the existence of absolute motion is incompatible with the many successes of both the
Special Theory of Relativity and of the General Theory of Relativity.  Both of these beliefs are demonstrably
false, see \cite{NovaBook,GQF}. 

The results in this section suggest, just as for Newtonian
gravity, that the Einstein General Relativity is nothing more than the dynamical equations for a velocity flow field 
${\bf v}({\bf r },t)$, but that both are not the best such theory, as both missed the `dark matter' dynamical
effect, and also the absolute motion effect, as now manifested on the RHS of  (\ref{eqn:f3vorticitya}) and
(\ref{eqn:f3vorticityb}).  Hence  the non-flat spacetime\index{spacetime} construct appears to be merely an
unnecessary  artifact of the Einstein measurement protocol, which in turn was motivated by the mis-reporting of
the results of the Michelson-Morley experiment
\cite{NovaBook,AMGE}. The putative successes of General Relativity should thus be considered as an incomplete
insight into  the fluid flow dynamics of the quantum foam system, rather than as any  confirmation of the validity
of the spacetime formalism, and it was this insight that in
\cite{NovaBook,GQF} led, in part, to the flow dynamics in (\ref{eqn:f3vorticitya}) and (\ref{eqn:f3vorticityb}).   
Nevertheless let us show that GR  reduces to a `flow' formalism for a restricted class of metrics, namely those
that  involve a velocity field.  To that end we  substitute the metric 
\begin{equation}
d\tau^2=g_{\mu\nu}dx^\mu dx^\nu=dt^2-\frac{1}{c^2}(d{\bf r}(t)-{\bf v}({\bf r}(t),t)dt)^2,
\label{eqn:PGmetric}\end{equation}
into (\ref{eqn:32})  using  (\ref{eqn:affine})  and (\ref{eqn:curvature}). This metric involves the arbitrary
time-dependent velocity field  ${\bf v}({\bf r},t)$.  The various components of the
Einstein tensor are then found to be
\begin{eqnarray}\label{eqn:G}
G_{00}&=&\sum_{i,j=1,2,3}v_i\mathcal{G}_{ij}
v_j-c^2\sum_{j=1,2,3}\mathcal{G}_{0j}v_j-c^2\sum_{i=1,2,3}v_i\mathcal{G}_{i0}+c^2\mathcal{G}_{00}, 
\nonumber\\ G_{i0}&=&-\sum_{j=1,2,3}\mathcal{G}_{ij}v_j+c^2\mathcal{G}_{i0},   \mbox{ \ \ \ \ } i=1,2,3.
\nonumber\\ G_{ij}&=&\mathcal{G}_{ij},   \mbox{ \ \ \ \ } i,j=1,2,3.
\end{eqnarray}
where the  $\mathcal{G}_{\mu\nu}$ are  given by
\begin{eqnarray}\label{eqn:GT}
\mathcal{G}_{00}&=&\frac{1}{2}((trD)^2-tr(D^2)), \nonumber\\
\mathcal{G}_{i0}&=&\mathcal{G}_{0i}=-\frac{1}{2}(\nabla\times(\nabla\times{\bf v}))_i,   \mbox{ \ \ \ \ }
i=1,2,3.\nonumber\\ 
\mathcal{G}_{ij}&=&
\frac{d}{dt}(D_{ij}-\delta_{ij}trD)+(D_{ij}-\frac{1}{2}\delta_{ij}trD)trD\nonumber\\ & &
-\frac{1}{2}\delta_{ij}tr(D^2)-(D\Omega-\Omega D)_{ij},  \mbox{ \ \ \ \ } i,j=1,2,3.
\end{eqnarray}
In vacuum, with $T_{\mu\nu}=0$, we find from (\ref{eqn:32}) and (\ref{eqn:G}) that $G_{\mu\nu}=0$ implies that  
$\mathcal{G}_{\mu\nu}=0$. This system of equations is thus very similar to the in-flow dynamics in
(\ref{eqn:f3vorticitya}) and (\ref{eqn:f3vorticityb}), except that in vacuum GR, for the  Panlev\'{e}-Gullstrand
metric, demands that
\begin{equation}
((trD)^2-tr(D^2))=0.
\label{eqg:DMGR}\end{equation}  
This simply corresponds to the fact that GR does not permit the `dark matter' effect, namely that
$\rho_{DM}=0,$ according to (\ref{eqn:DMdensity0}), and this happens because GR was forced to agree with Newtonian
gravity, in the appropriate limits, and that theory also has no such effect. As well in GR the energy-momentum
tensor
$T_{\mu\nu}$ is not permitted to make any reference to absolute linear motion of the matter; only  the relative
motion of matter or absolute rotational motion is permitted.

It is very significant to note that the above exposition of the GR formalism for  the  Panlev\'{e}-Gullstrand
metric is exact. Then taking the trace of the $\mathcal{G}_{ij}$ equation in (\ref{eqn:GT}) we obtain,
also exactly, and again using the identity in  (\ref{eqn:identity}), and in the case of zero vorticity,
and outside of matter so that
$T_{\mu\nu}=0$,
\begin{equation}
\frac{\partial }{\partial t}(\nabla.{\bf v})+\nabla.(({\bf
v}.{\bf \nabla}){\bf v})=0,
\label{eqn:f3vacuum}\end{equation}
which is the Newtonian `velocity field' formulation of Newtonian gravity outside of matter.  
This should have been expected as it corresponds to  the previous observation that  `Newtonian in-flow' velocity
field is exactly equivalent to the external Schwarzschild metric. So again we see  the extreme paucity of new
physics in the GR formalism:  all the key tests of GR are now seen to amount to a test {\it only} of $\delta
\tau[x]/\delta x^\mu = 0$,  when the in-flow field is given by  (\ref{eqn:GT}), and which
is nothing more than Newtonian gravity. Of course Newtonian gravity was itself merely based upon observations within
the solar system, and this was too special to have revealed key aspects of gravity. Hence, despite popular
opinion, the GR formalism is  based upon  very poor evidence. Indeed there is only one 
definitive confirmation of the GR formalism apart from  the misleading external-Schwarzschild metric cases, namely
the observed decay of  the binary pulsar orbital motion, for only in this case is the metric non-Schwarzschild, and
therefore non-Newtonian.  However the new theory of
gravity also leads to the decay of orbits, and on the grounds of dimensional analysis we would expect
comparable predictions.  So  GR is not unique in predicting orbital decay.

\section{  Apparent Invariance of c\label{section:apparentinvariance}}

The quantum foam induces actual dynamical time dilations and length contractions in agreement
with the Lorentz interpretation of special relativistic effects.  As a consequence of this  observers in 
uniform motion `through' the foam will on measurement  of the speed of light obtain always the same numerical
value  $c$, so long as they do not adjust their observational data to take account of these dynamical
effects.   So the special relativistic effects are very much an aspect of physical reality, but nevertheless
the absolute motion causing these effects is observable.  

To see this explicitly consider how various observers
$P, P^\prime,..$ moving with  different  speeds through the foam, might measure the speed of light.  They  each
acquire a standard rod  and an accompanying standardised clock. That means that these standard  rods  would
agree if they were brought together, and at rest with respect to the quantum foam they would all have length
$\Delta l_0$, and similarly for the clocks.    Observer $P$ and accompanying rod are both moving at  speed
$v_R$ relative to the quantum foam, with the rod longitudinal to that motion, for simplicity. P  then  measures
the time
$\Delta t_R$, with the clock at end $A$ of the rod,  for a light pulse to travel from  end $A$ to the other end
$B$  and back again to $A$. The  light  travels at speed $c$ relative to the quantum-foam. Let the time taken for
the light pulse to travel from
$A\rightarrow B$ be $t_{AB}$ and  from $B\rightarrow A$ be $t_{BA}$, as measured by a clock at rest with respect
to the quantum foam. The  length of the rod 
moving at speed
$v_R$ is contracted to 
\begin{equation}
\Delta l_R=\Delta l_0\sqrt{1-\frac{v_R^2}{c^2}}.
\label{eqn:c0}\end{equation}
In moving from  $A$ to $B$ the light must travel an extra  distance 
because the  end  $B$ travels a distance $v_Rt_{AB}$ in this time, thus the total distance that must be
traversed  is
\begin{equation}\label{eqn:c1}
ct_{AB}=\Delta l_R+v_Rt_{AB},
\end{equation}
Similarly on returning from $B$ to $A$ the light must travel the distance
\begin{equation}\label{eqn:c2}
ct_{BA}=\Delta l_R-v_Rt_{BA}.
\end{equation}
Hence the total travel time $\Delta t_0$ is
\begin{eqnarray}\label{eqn:c3}
\Delta t_0=t_{AB}+t_{BA}&=&\frac{\Delta l_R}{c-v_R}+\frac{\Delta l_R}{c+v_R}\\
&=&\frac{2\Delta l_0}{c\sqrt{1-\displaystyle\frac{v_R^2}{c^2}}}.
\end{eqnarray}
Because  of  the time dilation effect for the moving clock
\begin{equation}
\Delta t_R=\Delta t_0\sqrt{1-\displaystyle\frac{v_R^2}{c^2}}.
\label{eqn:c4}\end{equation}
Then for the moving observer the speed of light is defined as the distance the observer believes the light
travelled ($2\Delta l_0$) divided by the travel time according to the accompanying clock ($\Delta t_R$), namely 
$2\Delta l_0/\Delta t_R =c$.  So the speed $v_R$ of the observer through the quantum foam  is not revealed by this
procedure, and the observer is erroneously led to the conclusion that the speed of light is always c. 
This {\it invariance of c} follows from two or more observers in manifest relative motion all obtaining the same
speed c by this procedure. Despite this failure  this special effect is actually the basis of the spacetime
 measurement protocol. That this protocol is blind to the absolute motion has led to enormous confusion
within physics. However it is possible to overcome the `blindness' of this procedure and to manifestly reveal
an observer's   absolute velocity of  motion $v_R$. Several demonstrated techniques were given in \cite{NovaBook}.

\section{  The Lorentz Transformation\label{section:thelorentz}}

Here we show that the real dynamical effects of absolute moton results in certain special observational data
being related by the Lorentz transformation. This involves the use of the  radar measurement protocol
for acquiring observational space and time data of distant events, and subsequently  displaying that data in a
spacetime construct. In this protocol the observer records the time of emission and reception of radar pulses ($t_r
> t_e$) travelling through the space of quantum foam, and then retrospectively assigns the time and distance of a
distant event
$B$ according to (ignoring directional information for simplicity) 
\begin{equation}T_B=\frac{1}{2}(t_r+t_e), \mbox{\ \ \ }
D_B=\frac{c}{2}(t_r-t_e),\label{eqn:25}\end{equation}
  where each observer is now using the same numerical value of $c$.
 The event $B$ is then plotted as a point in 
an individual  geometrical construct by each  observer,  known as a spacetime record, with coordinates $(D_B,T_B)$. This
is no different to a historian recording events according to  some agreed protocol.  
  We now show that because of this
protocol and the quantum foam dynamical effects, observers will discover on comparing their
historical records of the same events that the expression
\begin{equation}
 \tau_{AB}^2 =   T_{AB}^2- \frac{1}{c^2} D_{AB}^2,
\label{eqn:26}\end{equation}
is an invariant, where $T_{AB}=T_A-T_B$ and $D_{AB}=D_A-D_B$ are the differences in times and distances
assigned to events $A$ and
$B$ using the above measurement protocol (\ref{eqn:25}), so long as both are sufficiently small
compared with the scale of inhomogeneities  in the velocity field. 

\begin{figure}[ht]
\vspace{-20mm}
\hspace{10mm}
\setlength{\unitlength}{2.0mm}
\hspace{30mm}\begin{picture}(40,45)
\thicklines
\put(-2,-2){{\bf $A$}}
\put(+1,30){{\bf $P(v_0=0)$}}
\put(16,14){\bf $B$ $(t^\prime_B)$}
\put(25,-3){\bf $D$}
\put(15,-3){\bf $D_B$}
\put(16,-0.5){\line(0,1){0.9}}
\put(-5,15){\bf $T$}
\put(26,24){$ P^\prime(v^\prime_0$)}

\put(0,0){\line(1,0){35}}
\put(0,0){\line(0,1){35}}
\put(0,35){\line(1,0){35}}
\put(35,0){\line(0,1){35}}
\put(0,0){\line(1,1){25}}

\put(0,8){\vector(2,1){16}}
\put(16,16){\vector(-2,1){16}}

\put(-3,8){\bf $t_e$}\put(-0.5,8){\line(1,0){0.9}}
\put(-3,16){\bf $T_B$}\put(-0.5,16){\line(1,0){0.9}}
\put(-3,24){\bf $t_r$}\put(-0.5,24){\line(1,0){0.9}}
\put(6,12){$\gamma$}
\put(6,22){$\gamma$}

\end{picture}
\vspace{5mm}
\caption{\small  Here $T-D$ is the spacetime construct (from  the  measurement protocol) of a special observer
$P$ {\it at rest} wrt the quantum foam, so that $v_0=0$.  Observer $P^\prime$ is moving with speed
$v^\prime_0$ as determined by observer $P$, and therefore with speed $v^\prime_R=v^\prime_0$ wrt the quantum foam. Two light
pulses are shown, each travelling at speed $c$ wrt both $P$ and the quantum foam.   Event
$A$ is when the observers pass, and is also used to define zero time  for each for
convenience. }\label{fig:spacetime1}
\end{figure}
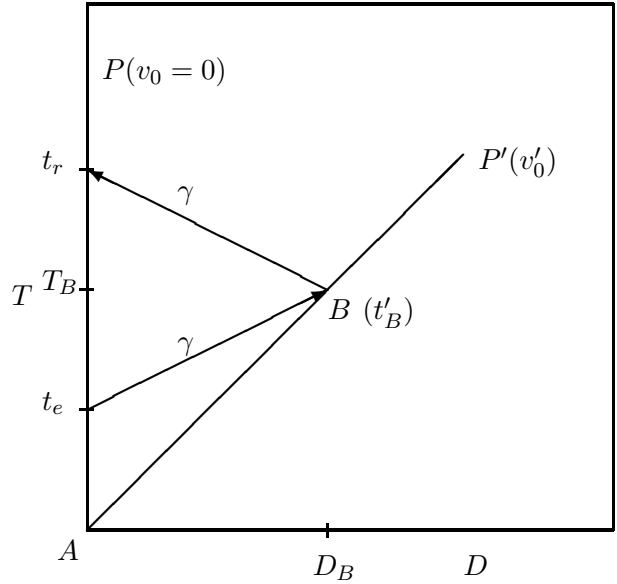

To confirm the invariant  nature of the construct in   (\ref{eqn:26}) one must pay careful attention to
observational times as distinct from protocol times and distances, and this must be done separately for each
observer.  This can be tedious.  We now  demonstrate this for the situation illustrated in
Fig.\ref{fig:spacetime1}. 

 By definition  the speed of
$P^\prime$ according to
$P$ is
$v_0^\prime =D_B/T_B$ and so
$v_R^\prime=v^\prime_0$,  where 
$T_B$ and $D_B$ are the protocol time and distance for event $B$ for observer $P$ according to
(\ref{eqn:25}).  Then using (\ref{eqn:26})  $P$ would find that
$(\tau^P_{AB})^2=T_{B}^2-\frac{1}{c^2}D_B^2$ since both
$T_A=0$ and $D_A$=0, and whence $(\tau^{P}_{AB})^2=(1-\frac{v_R^{\prime 2}}{c^2})T_B^2=(t^\prime_B)^2$ where
the last equality follows from the time dilation effect on the $P^\prime$ clock, since $t^\prime_B$ is the time
of event
$B$ according to that clock. Then $T_B$ is also the time that $P^\prime$  would compute for event $B$ when
correcting for the time-dilation effect, as the speed $v^\prime_R$ of $P^\prime$ through the quantum foam is
observable by $P^\prime$.  Then $T_B$ is the `common time' for event $B$ assigned by both
observers. For
$P^\prime$ we obtain  directly, also from  (\ref{eqn:25}) and (\ref{eqn:26}), that
$(\tau^{P'}_{AB})^2=(T_B^\prime)^2-\frac{1}{c^2}(D^\prime_B)^2=(t^\prime_B)^2$, as $D^\prime_B=0$  and
$T_B^\prime=t^\prime_B$. Whence for this situation
\begin{equation}
(\tau^{P}_{AB})^2=(\tau^{P'}_{AB})^2,
\label{eqn:invariant1}
\end{equation} and so the
 construction  (\ref{eqn:26})  is an invariant.  

While so far we have only established the invariance of the construct  (\ref{eqn:26}) when one of the
observers is at rest wrt to the quantum foam, it follows that for two observers $P^\prime$ and
$P^{\prime\prime}$ both in motion wrt the quantum foam it follows that they also agree on the invariance
of (\ref{eqn:26}).  This is easily seen by using the intermediate step of  a stationary observer $P$:
\begin{equation}
(\tau^{P'}_{AB})^2=(\tau^{P}_{AB})^2=(\tau^{P''}_{AB})^2.
\label{eqn:invariant2}
\end{equation}
Hence the measurement protocol and Lorentzian effects result in the construction in (\ref{eqn:26})  being indeed an
invariant in general.  This  is  a remarkable and subtle result.  For Einstein this invariance was a
fundamental assumption, but here it is a derived result, but one which is nevertheless deeply misleading.
Explicitly indicating  small quantities  by $\Delta$ prefixes, and on comparing records retrospectively, an
ensemble of nearby observers  agree on the invariant
\begin{equation}
\Delta \tau^2=\Delta T^2-\frac{1}{c^2}\Delta D^2,
\label{eqn:31}\end{equation} 
for any two nearby events.  This implies that their individual patches of spacetime records may be mapped one
into the other merely by a change of coordinates, and that collectively the spacetime patches  of all may
be represented by one pseudo-Riemannian manifold, where the choice of coordinates for this manifold is
arbitrary, and we finally arrive at the invariant 
\begin{equation}
\Delta\tau^2=g_{\mu\nu}(x)\Delta x^\mu \Delta x^\nu,
\label{eqn:inv}\end{equation} 
with $x^\mu=\{T,D_1,D_2,D_3\}$.  For flat metrics (\ref{eqn:inv}) is invariant under the well known Lorentz
transformation,
\begin{equation}\label{eqn:LT1}
x^\mu=L({\bf v})^\mu_{\mbox{\ }\nu} x^{\prime\nu},
\end{equation}
where, for motion only in the x-direction,
\begin{eqnarray}\label{eqn:LT2}
x&=&\gamma(x^\prime-\beta c t^\prime) \nonumber \\
ct&=&\gamma(ct^\prime-\beta x^\prime) \nonumber \\
y&=&y^\prime \nonumber \\
z&=&z^\prime
\end{eqnarray}
where  $\beta=v/c$ and $\gamma=1/\sqrt{1-\beta^2}$.
Here, in general, ${\bf v}$ is the relative velocity of the two observers,  determined by using the measurement
protocol.  The special feature of this mapping between the observer's spacetime records is that it does {\it
not} involve the absolute velocity of either observer relative to the quantum-foam substratum - their absolute
velocities. This feature was responsible, essentially,  for  the Einsetin assumption about `c' being invariant
for all observers in uniform motion. This feature has caused enormous confusion in physics. It erroneously
suggests that absolute motion is incompatible with relativistic effects - that the observation of absolute motion
must be in conflict with the observation of relativistic effects.  For that reason reports of the ongoing
detection of absolute motion has been banned in physics for nearly 100 years.    However to the contrary absolute
motion {\it and} special relativistic effects are both needed to understand and analyse the extensive experimental
data reported in
\cite{AMGE}. The key insight is that absolute motion dynamically causes the time dilation
and length contraction effects. Without absolute motion there would be no special relativistic effects.   This
insight runs counter to nearly 100 years of conventional wisdom within physics.

\section{ The `Dark Matter' Effect \label{section:thedark}}

 Because of the  $\alpha$ dependent term    (\ref{eqn:f3vorticitya}) and
(\ref{eqn:f3vorticityb}) would predict that the Newtonian inverse square law
would not be applicable  to systems such as spiral galaxies. A detailed
analysis and comparison with experimental and observational data is given in
\cite{alpha,NovaDM}.   Of course attempts to retain this law, despite its  
manifest failure,  has   led to the  spurious introduction of  the notion of `dark
matter' within spiral galaxies, and also at larger scales.  Here we merely note
the basic idea that the `dark matter' effect is indeed a dynamical effect of
space itself.  From 
\begin{equation}\label{eqn:ga}
{\bf g}=\frac{1}{2}\nabla({\bf v}^2)+\frac{\partial {\bf v}}{\partial
t},\end{equation} 
which is (\ref{eqn:f2}) for irrotational flow and in the limit ${\bf}_R\rightarrow 0$, and in that case we
see that (\ref{eqn:f3vorticitya}) and (\ref{eqn:f3vorticityb}) give (\ref{eqn:f3extend}), which implies
that
\begin{equation}\label{eqn:g2}
\nabla.{\bf g}=-4\pi G\rho-C({\bf v}),
\end{equation}
and taking running time averages to account for turbulence
\begin{equation}\label{eqn:g3}
\nabla.\!\!<\!\!{\bf g}\!\!>=-4\pi G\rho-<\!\!C({\bf v})\!\!>,
\end{equation}
and writing  the extra term as $<\!\!C({\bf v})\!\!>=4\pi G \rho_{DM}$ we see that  $\rho_{DM}$,
introduced in (\ref{eqn:DMdensity0}), would act as an effective matter density, and it is now clear,
as shown in \cite{alpha,NovaDM}, that it is the consequences of this term which have been misinterpreted
as `dark matter'. We thus see that this effect is actually the consequence of quantum foam effects within
the new proposed dynamics for gravity, and which becomes apparent particularly in spiral galaxies.  Note
that (\ref{eqn:f3extend}) for $\nabla\times{\bf v}=\bf{0}$ becomes an equation for the velocity potential
$u({\bf r},t)$, 
\begin{equation}
\nabla^2\left(\frac{\partial u}{\partial t}+\frac{1}{2}(\nabla u)^2\right)=-4\pi G\rho-C(\nabla u({\bf r})).
\label{eqn:ueqn}\end{equation}
Then noting (\ref{eqn:deltafnidentity})
 we see that (\ref{eqn:ueqn}) has the non-linear integro-differential equation form
\begin{equation}\label{eqn:ueqn2}
\frac{\partial u({\bf r},t)}{\partial t}=-\frac{1}{2}(\nabla u({\bf r},t))^2+\frac{1}{4\pi}\int d^3
r^\prime\frac{C(\nabla u({\bf r}^\prime,t))}{|{\bf r}-{\bf r}^\prime|}-\Phi({\bf r},t),
\end{equation}
where $\Phi$ is the Newtonian gravitational potential
\begin{equation}\label{eqn:Phieqn}
\Phi({\bf r},t)=-G\int d^3 r^\prime\frac{\rho({\bf r}^\prime,t)}{|{\bf r}-{\bf r}^\prime|}.
\end{equation}
Hence the  $\Phi$  field acts as the source term for  the velocity potential. Note that in the Newtonian
theory of gravity one has the choice of using either the acceleration field ${\bf g}$ or the velocity field
${\bf v}$. However in the new theory of gravity this choice is no longer available: the fundamental
dynamical degree of freedom is necessarily the ${\bf v}$ field, again because of the presence of the $C({\bf
v})$ term, which obviously cannot be written in terms of ${\bf g}$.  If we were to ignore time-dependent behaviour
(\ref{eqn:ueqn2}) gives
\begin{equation}\label{eqn:veqnagain}
|{\bf v}({\bf r})|^2=\frac{2}{4\pi}\int d^3
r^\prime\frac{C({\bf v}({\bf r}^\prime))}{|{\bf r}-{\bf r}^\prime|}-2\Phi({\bf r}).
\end{equation}
This non-linear equation clearly cannot be solved for ${\bf v}({\bf r})$ as its direction is not
specified.  This form makes it clear that we should expect gravitational waves, but certainly not waves travelling at
the speed of light as $c$ does not appear in (\ref{eqn:ueqn2}). Note that (\ref{eqn:ueqn2}) involves
`action-at-a-distance' effects, as there is no time-delay in the denominators. This was a feature of Newton's
original theory of gravity.  Here it is understood to be caused by the underlying quantum-foam dynamics (QHFT) which
reaches this classical `flow' description by ongoing non-local and instantaneous wavefunctional collapses, as
discussed in
\cite{NovaBook}. Contrary to popular belief even GR has this `action-at-a-distance' feature, as the reformulation of
GR via the Panlev\'{e}-Gullstrand metric  leads also  to an equation of the form in   (\ref{eqn:ueqn}), but
with the $C({\bf v})$ term absent.

\section{ Observations of  Absolute Motion and Gravitational In-Flows \label{section:observationsof}}

An extensive analysis of numerous experimental observations of absolute motion has been reported in
\cite{CK,AMGE}. Absolute motion is motion relative to space itself.  It turns out that Michelson and
Morley \cite{MM} in their historic experiment of 1887 did detect absolute motion, but rejected their own
findings because, using a flawed model for the operation of the interferometer, the analysis of their
data led to a speed of some 8 km/s, which  was less than the 30 km/s orbital speed of the earth. Key
aspects missing from their theory were, in addition to the known geometrical effect of the different path
lengths when in translation,  the Fitzgerald-Lorentz contraction and the effect of the gas in slowing the
speed of the light (the refractive index effect). The data to the contrary clearly indicated evidence of
absolute motion and, furthermore, that the theory for the operation of the Michelson interferometer was
not adequate.  Rather than reaching this conclusion Michelson and Morley came to the incorrect conclusion
that their results amounted to the failure to detect absolute motion, even though the data clearly showed
a signal with the expected signature, namely an $180^0$ period  on rotating the interferometer.   This
had an enormous impact on the development of physics, for as is well known Einstein adopted the absence 
of absolute motion effects as one of his fundamental assumptions.  By the time Miller
\cite{Miller2} had finally figured out how to use and properly analyse data from his Michelson interferometer absolute motion
had become a forbidden concept within physics, as it still is at present.  The  experimental observations  by Miller
and others of absolute motion has continued to be scorned and rejected by the physics community.  Fortunately
as well as revealing absolute motion the experimental data also reveals evidence in support of a new theory of
gravity.    

In ref.\cite{CK,AMGE} the analysis of data from six experiments
demonstrated that absolute motion  relative to space  has been observed  by Michelson
and Morley \cite{MM}, Miller \cite{Miller2}, Illingworth  \cite{Illingworth}, Jaseja {\it et al} \cite{Jaseja}, Torr and
Kolen \cite{Torr}, and by  DeWitte \cite{DeWitte}, contrary to common belief within physics that absolute motion has never
been observed.   The  Dayton Miller   also reveals via the analysis in \cite{AMGE}, the in-flow of
space past the earth into the sun.  The direction of the cosmic absolute velocity is found to be different to that
of the CMB due to the in-flow into the Milky Way and the local galactic cluster, see
Sect.\ref{section:galacticinflow}.  The Miller experimental data also suggests that the in-flow manifests
turbulence, as does the DeWitte data, which amounts to the observation of a gravitational wave phenomena.

The extensive experimental data shows that absolute motion is consistent with relativistic
effects. Indeed relativistic effects are caused by dynamical effects associated with absolute
motion, as proposed by Lorentz, and relativistic effects are required in understanding the gas-mode Michelson interferometer
experiments. The Lorentz transformation is seen to be a consequence of absolute motion dynamics.  Vacuum Michelson interferometer
experiments or its equivalent
\cite{vacuum, KT, BH, Muller, NewVacuum} cannot detect absolute motion, but their null results do
support this interpretation and form a part of the experimental predictions of the new physics.

\section{ The Velocity Superposition Principle\label{section:thesuperposition}}

Despite being non-linear (\ref{eqn:f3vorticitya})-(\ref{eqn:f3vorticityb}) possess an approximate
superposition principle\footnote{This section has been considerably changed since the September 2003
version of this paper.}, which explains  why the existence of absolute motion and as well the presence of
the $C({\bf v})$ term appears to have escaped attention in the case of gravitational experiments and
observations near the earth, despite the fact, in the case of the $C({\bf v})$ term, that the presence of
the earth breaks the spherical symmetry of the matter distribution of   the sun. 

First note that  in analysing (\ref{eqn:f3vorticitya})-(\ref{eqn:f3vorticityb}) we need to recognise
two distinct effects: (i) the effect of a change of description of the flow when changing between 
observers, and (ii) the effects of absolute motion of the matter with respect to the quantum foam
substratum. Whether the matter is at rest or in absolute motion with respect to this substratum does
have a dynamical effect, albeit  very small.  While the Newtonian theory and GR both offer an account of the
first effect, and different accounts at that, neither have the second dynamical effect, as this is a unique
feature of the new theory of gravity.  Let us  consider the first effect, as this is somewhat standard. It
basically comes down to noting that under a change of observer (\ref{eqn:f3vorticitya})-(\ref{eqn:f3vorticityb})
transform covariantly under a Galilean transformation.
Suppose that according to one observer $O$ the matter density is specified by  a form $\rho_O({\bf r},t)$,
and that (\ref{eqn:f3vorticitya})-(\ref{eqn:f3vorticityb}) has a solution
${\bf v}_O({\bf r},t)$, and then  with acceleration ${\bf g}_O({\bf r},t)$ given by (\ref{eqn:G6})\footnote{Note
that here and in the following, except where indicated, the subscripts  are $O$ and not $0$.}.  Then
 for  another observer $O^\prime$ (and for  simplicity we assume that the observers  use coordinate axes that
have the same orientation, and that at time $t=0$ they coincide), moving with uniform velocity ${\bf V}$ relative
to observer $O$,  observer $O^\prime$ describes the matter density with the form
$\rho_{O^\prime}({\bf r},t)=  \rho_O({\bf r}+{\bf V}t,t)$. Then, as we now show, the corresponding  solution to
(\ref{eqn:f3vorticitya})-(\ref{eqn:f3vorticityb}) for $O^\prime$ is {\it exactly} 
 \begin{equation}\label{eqn:Vsum}
{\bf v}_{O^\prime}({\bf r},t)={\bf v}_O({\bf r}+{\bf V}t,t)-{\bf V}.
\end{equation}
This is easily established by substitution of (\ref{eqn:Vsum}) into
(\ref{eqn:f3vorticitya})-(\ref{eqn:f3vorticityb}), and noting that the LHS leads to  a RHS where the density has
the different form noted above, but that ${\bf v}_R$ is {\it invariant} under this change of observer, for each
observer agrees on the absolute velocity of each piece of matter with respect to the local quantum foam.  Under
the change of observers, from $O$ to $O^\prime$,  (\ref{eqn:Vsum}) gives
\begin{equation} D_{ij}({\bf r},t) \rightarrow D_{ij}({\bf r+V}t,t) \mbox{\ \ and \ \ } 
 \Omega_{ij}({\bf r},t) \rightarrow \Omega_{ij}({\bf r+V}t,t).
\end{equation}
Then for the total or Euler fluid derivative in (\ref{eqn:f3vorticitya}) we have for observer $O^\prime$ 
\begin{eqnarray}\label{eqn:covariant}
\frac{d D_{ij}({\bf r+V}t,t)}{dt} &\equiv&
\frac{\partial D_{ij}({\bf r+V}t,t)}{\partial t}+({\bf v}_O({\bf r}+{\bf V}t,t)-{\bf V}).{\bf \nabla}D_{ij}({\bf
r+V}t,t),\nonumber\\ &=&
\left.\frac{\partial D_{ij}({\bf r+V}t^\prime,t)}{\partial t^\prime}\right|_{t^\prime\rightarrow t}+
\left.\frac{\partial D_{ij}({\bf r+V}t,t^{\prime\prime}))}{\partial
t^{\prime\prime}}\right|_{t^{\prime\prime}\rightarrow t} + \nonumber \\&&\mbox{\ \ \ \ \ \ \ \ \ \ \ \ \ \ \ \  }
({\bf v}_O({\bf r}+{\bf V}t,t)-{\bf V}).{\bf
\nabla}D_{ij}({\bf r+V}t,t),\nonumber\\ &=&
({\bf V}.\nabla)D_{ij}({\bf r+V}t,t)+
\left.\frac{\partial D_{ij}({\bf r+V}t,t^{\prime\prime}))}{\partial
t^{\prime\prime}}\right|_{t^{\prime\prime}\rightarrow t} + \nonumber \\&&\mbox{\ \ \ \ \ \ \ \ \ \ \ \ \ \ \ \  }
({\bf v}_O({\bf r}+{\bf V}t,t)-{\bf V}).{\bf \nabla}D_{ij}({\bf r+V}t,t),\nonumber\\
&=&
\left.\frac{\partial D_{ij}({\bf r+V}t,t^{\prime\prime}))}{\partial
t^{\prime\prime}}\right|_{t^{\prime\prime}\rightarrow t} + 
{\bf v}_O({\bf r}+{\bf V}t,t).{\bf \nabla}D_{ij}({\bf r+V}t,t),\nonumber\\
&=&\left.\frac{d D_{ij}({\bf r},t))}{dt}\right|_{\displaystyle{{\bf r}\rightarrow {\bf r}+{\bf V}t}}
\end{eqnarray} 
as there is a key cancellation of two terms in (\ref{eqn:covariant}).  Clearly then all the terms on the LHS of 
(\ref{eqn:f3vorticitya})-(\ref{eqn:f3vorticityb}) have the same  transformation property. Then, finally, from the
form of the LHS,   both equations give the density dependent RHS, but which now  involves  the form
$  \left.\rho_O({\bf r},t)\right|_{\displaystyle{{\bf r}\rightarrow {\bf r}+{\bf V}t}}\mbox{\  }$, and this is
simply $\rho_{O^\prime}({\bf r},t)$ given above. 
If the observers coordinate axes do not have the same orientation then a time-independent orthogonal similarity
transformation $D \rightarrow SDS^T$,  $\Omega \rightarrow S\Omega S^T$, and ${ v}^i_R\rightarrow 
\sum_jS_{ij}v^j_R$ arises as well. Hence the description of the flow dynamics for observers in uniform relative
motion is Galilean covariant. While the transformation rule for the Euler derivative is not a new result, there
are some subtleties in the analysis, as seen above.  The subtlety arises because the change of coordinate
variables  necessarily introduces a time dependence in the observer descriptions, even if the flow is
inherently stationary.   Finally, using an analogous argument to that in (\ref{eqn:covariant}),  we see explicitly
that the acceleration is also Galilean covariant under the above change of observer, ${\bf g}({\bf
r},t)\rightarrow{\bf g}({\bf r}+{\bf V}t,t)$ (in the case of observer axes with the same orientation). This
simply asserts that all observers actually agree on the gravitational  acceleration. 

We now come to item (ii)  above, namely the more subtle but experimentally significant {\it approximate}
velocity superposition principle.  This approximate effect relates to the change in the form of the solutions of
(\ref{eqn:f3vorticitya})-(\ref{eqn:f3vorticityb}) when the matter density is in motion, as a whole, with respect
to the quantum-foam substratum, as compared to the solutions when the matter is, as a whole, at rest.   Already
even these descriptions involve a subtlety. Consider the case when a star, say, is `at rest' with respect to the
substratum. Then the flow dynamics in (\ref{eqn:f3vorticitya})-(\ref{eqn:f3vorticityb}) will lead to a position
and time dependent flow solution ${\bf v}({\bf r},t)$. But that flow leads to a position
and time dependent  ${\bf v}_R({\bf r},t)={\bf v}_0({\bf r},t)-{\bf v}({\bf r},t)$ on the RHS of
(\ref{eqn:f3vorticitya})-(\ref{eqn:f3vorticityb}), where  ${\bf v}_0({\bf r},t)$  is the velocity of the matter at
position ${\bf r}$ and time $t$ according to some specific observer's frame of reference\footnote{Here the
subscript is $0$ and not an $O$.  ${\bf v}_R({\bf r},t)$ was defined in (\ref{eqn:18}). For matter described by a
density distribution it is appropriate to introduce the field ${\bf v}_0({\bf r},t)$. }. Hence the description of
the matter being `at rest' or `in motion' relative to the substratum  is far from simple.  In general, with
time-dependent flows, none of the matter will ever be `at rest' with respect to the substratum, and this
description is covariant under a change of observer.   In the case of a well isolated star exisitng in a
non-turbulent substratum we could give the terms `at rest as a whole' or `moving as whole' a well defined meaning
by deciding how the star as a whole, considered as a rigid body,  was moving relative to the more distant
unperturbed substratum.  Despite these complexities the solutions of
(\ref{eqn:f3vorticitya})-(\ref{eqn:f3vorticityb}) have, under certain conditions, an approximate dynamical
velocity superposition princple, namely that if the flow is ${\bf v}_O({\bf r},t)$, according to some observer
$O$, when the matter is `at rest as a whole',  then when the same matter density is `moving as a whole' with
velocity ${\bf V}$, then the  flow solution  becomes {\it approximately} ${\bf v}_O^\prime({\bf r},t)\approx{\bf
v}_O({\bf r}-{\bf V}t,t)$, according to the same observer. Then if we want the flow field for the observer
$O^\prime$ travelling at the same velocity as the star relative to the quantum foam, then using the {\it exact}
Galilean  transformation in (\ref{eqn:Vsum}) in conjunction with ${\bf r}\rightarrow{\bf r}+{\bf V}t $, we obtain
${\bf v}_{O^\prime}^\prime({\bf r},t)\approx{\bf v}_O({\bf r},t)-{\bf V}$. That is the observer moving with the
star sees approximately the vector sum of  the in-flow field that the star would have when `at rest as a whole'
and the `flow' $-{\bf V}$ caused by the star and observer moving through the substratum with velocity ${\bf V}$
that the observer would detect in the absence of the star.  To be more explicit about the signs we note that if we
move in a southerly direction, say, against the substratum, then the substratum appears to be flowing though or
past us in a northerly direction.  To see why, dynamically, the above velocity superposition approximation is valid
note that the effect of absolute motion of matter forming the star, with respect to the local substratum, as
specified by   
${\bf v}_R({\bf r},t)$, arises at order $(v_R/c)^2$, and also with strength $G$, which together makes the RHS of
(\ref{eqn:f3vorticitya}) very insensitive to the absolute motion of the matter, and with a similar argument for
the vorticity in  (\ref{eqn:f3vorticityb}).  So only at the centre of stars, for example, where $v_R$ does become
large because of the gravitational attractor effect ({\it viz} the black hole effect), see  \cite{NovaDM}, will
this approximate superposition principle breakdown. It then follows that the derived acceleration field ${\bf
g}({\bf r},t)$ is also approximately unchanged, up to a Galilean transformation corresponding to an observer
dependent change of description, by the `motion as a whole' of the matter through the substratum.  

Hence the gravitational acceleration, say of the earth, will
only be affected,  by its observed absolute  motion, at order  $(v_R/c)^2$, and this has the approximate value of 
$10^{-6}$.   Note that this approximate dynamical velocity superposition principle is completely different to what
would occur if a solid object were moving through a normal classical fluid, for in this case we obviously would
not see any such superposition effect.  For  earth based  gravitational phenomena the motion of the earth takes
place within the inhomogeneous velocity  in-flow  towards the sun, and the velocity superposition  is only
approximately valid  as now
${\bf V}\rightarrow {\bf V}({\bf r},t)$ and no longer corresponds to uniform translation, and  manifests
turbulence.  To be a valid approximation the inhomogeneity of ${\bf V}({\bf r},t)$ must be much smaller than that
of  ${\bf v}({\bf r}-{\bf V}t,t)$, which it is, as the earth's centripetal acceleration about the sun is
approximately 1/1000 that of the earth's gravitational acceleration at the surface of the earth.  Of course the
inhomogeneity of the flow towards the sun is what keeps the earth in its orbit about the sun, and indeed that
component of the flow is now understood to be present in the  Miller absolute motion data \cite{AMGE}. 
So the validity of this velocity superposition approximation  demonstrates  that the detection of a cosmic absolute
motion and of  the in-flow theory of gravity are consistent with the older methods of computing gravitational
forces. This is why both the presence of the $C({\bf v})$ term, the in-flow and the absolute motion  have gone
almost unnoticed in earth based gravitational experiments, except for the so-called various `anomalies'; see
Sect.\ref{section:measurementsofG}.

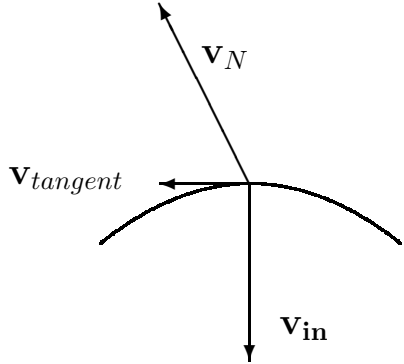
\begin{figure}[ht]
\vspace{20mm}
\hspace{60mm}\setlength{\unitlength}{0.8mm}
\begin{picture}(0,20)
\thicklines
\put(25,10){\vector(-1,0){15}}
\put(25,10){\vector(0,-1){29.5}}
\put(25,10){\vector(-1,2){15}}
\qbezier(0,0)(25,20)(50,0)
\put(30,-15){\Large $\bf v_{in}$}
\put(-15,10){\Large ${\bf v}_{tangent}$}
\put(17,30){\Large ${\bf v}_{N}$}
\end{picture}
\vspace{20mm} 
\caption{\small  Orbit of earth about the sun with tangential orbital velocity
${\bf v}_{tangent}$ and quantum-foam in-flow velocity  ${\bf v}_{in}$. Then ${\bf v}_{N}={\bf
v}_{tangent}-{\bf v}_{in}$ is the velocity of the earth relative to the quantum foam, after subtracting
${\bf v}_{cosmic}$. Alternatively this diagram also represents the various velocities associated with a
satellite in orbit about the earth.}
\label{fig:orbit}\end{figure}

This approximate velocity superposition principle is the key to understanding the operation of the
 earth based detections of absolute motion \cite{AMGE}.  There are four  velocities that contribute to the total
 velocity of an observer through space:
\begin{equation}\label{eqn:QG6b}
{\bf v} \approx {\bf v}_{cosmic} +{\bf v }_{tangent} -{\bf v}_{in}-{\bf v}_E.
\end{equation}
Here ${\bf v}_{cosmic}$ is the velocity of the solar system through space, while the other three are local 
 effects: (i) ${\bf v }_{tangent}$ is the tangential orbital velocity of the earth about the sun,  
(ii) ${\bf v}_{in}$ is  a quantum-foam radial in-flow   past the earth towards the
sun, and (iii) the corresponding quantum-foam in-flow into the
earth is ${\bf v}_E$ and makes no contribution to a horizontally operated   interferometer. In constructing
(\ref{eqn:QG6b}) we have asumed the validity of the velocity superposition principle, namely that ${\bf
v}_{cosmic}$, ${\bf v}_{in}$ and ${\bf v}_E$ may be approximately combined vectorially.  The minus signs in
(\ref{eqn:QG6b}) arise because, for example, the in-flow towards the sun requires the earth to have an outward
directed velocity component against that in-flow in order to maintain a fixed distance from the sun, as shown in
Fig.\ref{fig:orbit}.   For circular orbits   
$v_{tangent}$  and  $v_{in}$ are given by  
\begin{eqnarray}\label{eqn:QG7}
v_{tangent}&=&\sqrt{\displaystyle{\frac{GM}{R}}},\\  
v_{in}&=&\sqrt{\displaystyle{\frac{2GM}{R}}},\end{eqnarray}
while the net speed $v_N$ of the earth from the vector sum   ${\bf v}_N={\bf v}_{tangent}-{\bf
v}_{in}$  is 
\begin{equation}\label{eqn:QG7b}
v_{N}=\sqrt{\displaystyle{\frac{3GM}{R}}},\end{equation}     
where $M$ is the mass of the sun, $R$ is the distance of the earth from the
sun, and $G$ is Newton's gravitational constant. $G$ is essentially a measure of the rate at
which matter effectively `dissipates' the quantum-foam. The gravitational acceleration of the earth towards
the sun arises from  inhomogeneities in the $v_{in}$  flow component.  These expressions give
$v_{tangent}=30$km/s,  $v_{in}=42.4$km/s and
$v_{N}=52$km/s. As discussed in  \cite{AMGE} $v_{in}$ is extractable from Miller's 1925/26 air-mode Michelson
interferometer experiment, and so provided the first confirmation of the new `flow' theory of gravity. As noted
above the velocity superposition approximation implies that the gravitational entrainment or `drag effects' are
very small. Nevertheless they are present and the Gravity Probe B satellite experiment was designed to detect
just  such a vorticity  effect.  Of course the magnitude of this vorticity effect predicted by the new theory of
gravity \cite{alpha} is much larger than that predicted by General Relativity, for in this theory only {\it
rotation} has an absolute meaning, whereas in the new theory both rotation and linear motion with respect to the
substratum have an absolute meaning, and generate observable vorticity effects. As well the absolute linear
velocity of the earth through the substratum also has a larger effect on the so-called geodetic precession of the
GP-B gyroscopes, than predicted by General Relativity.    

\section{ Gravitational In-Flow and the GPS\label{section:gps}}

We show here that the new in-flow theory of gravity and the observed absolute velocity of motion of
the solar system through space are   compatible with the operation of the Global Positioning System
(GPS), and that the new theory of gravity finally provides a theory for the operation of the GPS, that is, that
the account given by General Relativity was actually only fortuitously correct.  Given the developments above  this
turns out to be an almost  trivial exercise.  As usual in this system the effects of the sun and moon are
neglected. Various effects need to be included as the system relies upon extremely accurate atomic clocks in the
satellites forming the GPS constellation.  Within both the new theory and General Relativity these clocks are
effected by both their speed and  the gravitational effects of the earth. As well the orbits of these satellites
and the critical time delays of radio signals from the satellites need to be computed.  For the moment we assume
spherical symmetry for the earth. The effects of non-sphericity will be discussed below.  In General Relativity
the orbits and signalling time delays are determined by the use of the geodesic equation (\ref{eqn:33}) and the
Schwarzschild metric (\ref{eqn:SM}).  However these two equations are equivalent to the  orbital equation
(\ref{eqn:G8}) and the velocity field  (\ref{eqn:Vsum}), with a velocity ${\bf V}$ of absolute motion, and  with
the in-flow given by (\ref{eqn:vfield}), noting the result in Sect.\ref{section:thesuperposition}. For EM
signalling the elapsed time in (\ref{eqn:f4}) requires careful treatment.  Hence the two systems are 
mathematically completely equivalent:  the computations within the new system may most easily be considered by
relating them to the mathematically equivalent General Relativity formalism.  
We can also see this by explicitly changing from the CMB frame to a non-rotating frame co-moving with the earth
by means of  the change of variables
\begin{eqnarray}\label{eqn:frame1}
{\bf r}&=&{\bf r}^\prime+{\bf V}t,  \\ \nonumber
t&=&t^\prime, \\  \nonumber
{\bf v}&=&{\bf v}^\prime+{\bf V},
\end{eqnarray}  
which lead to  the relationships of differentials
\begin{eqnarray}\label{eqn:frame2}
\nabla^\prime &=& \nabla, \\  \nonumber
\frac{\partial}{\partial t ^\prime}&=&\frac{\partial}{\partial t}+{\bf V}.\nabla
\end{eqnarray}
These expressions then lead to the demonstration of the covariance of
(\ref{eqn:f3vorticitya})-(\ref{eqn:f3vorticityb}). Then in the earth co-moving frame the absolute cosmic velocity
${\bf V}$ only appears on the RHS of these equations, and as noted in the previous section, has an extremely
small dynamical effect.  

   There are nevertheless two  differences between the two theories. One is their different
treatment of the non-sphericity of the earth  via the $C({\bf v})$ term, and the second difference is the effects
of the in-flow turbulence.  In the operation of the GPS the density $\rho({\bf r})$ of the earth is not used. 
Rather the gravitational potential $\Phi({\bf r})$ is determined observationally.  In the new gravity theory the
determination of such a gravitational potential via (\ref{eqn:f3vorticitya})-(\ref{eqn:f3vorticityb})  and 
$\Phi({\bf r})=-\frac{1}{2}{\bf v}^2({\bf r})$ would involve the extra  $C({\bf v})$ term.  Hence because of
this phenomenological treatment the effects of the  $C({\bf v})$ term are not checkable.   However the
gravitational wave effect is expected to affect the operation of the GPS, and the GPS constellation
would offer a worldwide network  which would enable the investigation of the spatial and temporal correlations
of these gravitational waves. 

There is also a significant interpretational difference between the two theories. For example in  General
Relativity the relativistic effects involve both the `special relativity' orbital speed  effect via time
dilations of the satellite clocks together with the General Relativity `gravitational potential energy' effect on
the satellite clocks. In the new theory there is only one effect, namely the time dilation effect produced by the
motion of the clocks through the quantum foam, and the speeds of these clocks involves the vector sum of the
orbital velocity and the velocity caused by the in-flow of the quantum foam into the earth.  This is illustrated by
Fig.\ref{fig:orbit}, where now the orbit refers to that of a satellite about the earth.

As well as providing a platform for studying the new gravitational waves, the GPS is already used for accurate time
transfers. But because General Relativity and the new theory of gravity are fundamentally different there will be
differences at higher orders in $v_R/c$.  A systematic study of these corrections should be undertaken with the
possibility that they will permit  the establishment of more accurate global time standards.  

\section{  Gravitational Anomalies \label{section:measurementsofG}}

{\it Gravitational anomalies} are those observed effects which are apparently inconsistent with either
 the Newtonian theory of gravity or General Relativity. The `dark matter' effect is in fact one such anomaly.  These anomalies  are in
the low speed/low mass regime appropriate  to the Newtonian theory,  but since  General Relativity was constructed to agree with 
Newtonian theory in this regime these anomalies are common to both theories.  Here we note that these 
anomalies appear to be explainable within the new theory of gravity, and are indicators of the failure of
Newtonian gravity and hence  General Relativity.

As noted in Sect.\ref{section:QFinflow} Newton's Inverse Square Law of Gravitation
may only be strictly valid in cases of spherical symmetry.  The  theory that gravitational effects
arise from inhomogeneities in the quantum foam flow  implies that there is no `universal law of
gravitation' because the inhomogeneities are determined by non-linear `fluid equations' and the
solutions  have no form which could be described by a `universal law'.  Fundamentally there is no
generic fluid flow behaviour. The Inverse Square Law is then only an approximation, with  large
deviations expected in the case of spiral galaxies. Nevertheless Newton's gravitational constant $G$
will have a definite value as it quantifies the effective rate at which matter dissipates the
information content of space.

\begin{figure}[t]
\hspace{20mm}\includegraphics[scale=1.4]{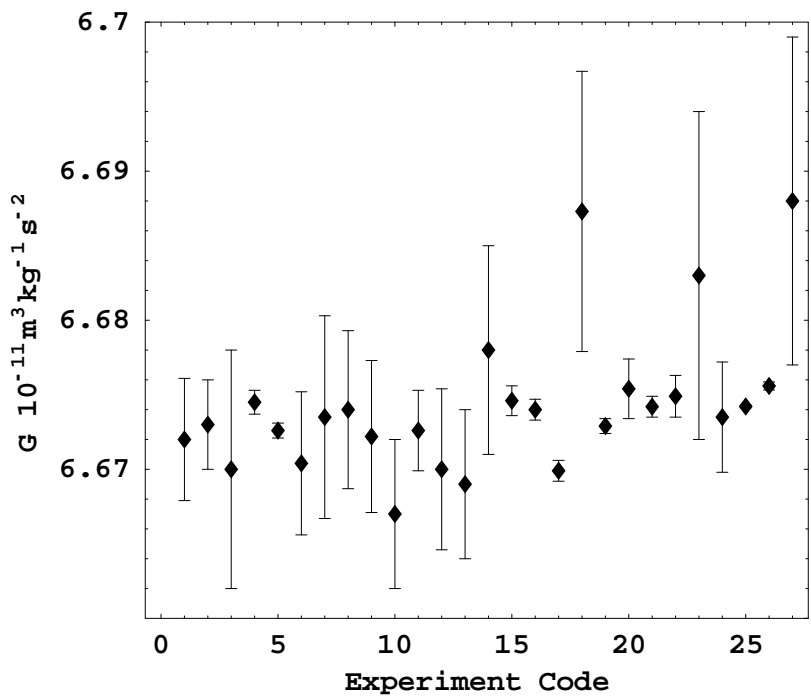}
\caption{\small{Results of precision measurements of $G$ published in the 
last sixty years in which the Newtonian theory was used to analyse the data.  These results show  the presence of
a  systematic effect not in the Newtonian theory. {\bf 1:}  Gaithersburg 1942 \cite{Gaithersburg42},
{\bf 2:}  Magny-les-Hameaux 1971 \cite{Magny-les-Hameaux}, 
{\bf 3:}  Budapest 1974   \cite{Budapest}, 
{\bf 4;}  Moscow 1979  \cite{Moscow79},
{\bf 5:}  Gaithersburg 1982 \cite{Gaithersburg82}, 
{\bf 6-19:}  Fribourg   Oct 84, Nov 84, Dec 84, Feb 85 \cite{Fribourg},
{\bf 10:}    Braunschweig 1987   \cite{Braunschweig87},
{\bf 11:}    Dye 3 Greenland   1995       \cite{Dye3},
{\bf 12:}    Gigerwald Lake 1994   \cite{Gigerwaldlake94}, 
{\bf 13-14:} Gigerwald lake19 95  112m, 88m     \cite{Gigerwaldlake95},
{\bf 15:}     Lower Hutt 1995    MSL  \cite{LowerHutt95}, 
{\bf 16:}  Los Alamos 1997 \cite{Los Alamos}, 
{\bf 17:}    Wuhan 1998  \cite{Wuhan},
{\bf 18:}    Boulder JILA 1998  \cite{Boulder}, 
{\bf 19:}    Moscow 1998  \cite{Moscow98}, 
{\bf 20:}    Zurich 1998  \cite{Zurich98}, 
{\bf 21:}     Lower Hutt MSL 1999   \cite{LowerHutt99}, 
{\bf 22:}    Zurich 1999   \cite{Zurich99}, 
{\bf 23:}    Sevres 1999  \cite{Sevres99}, 
{\bf 24:}   Wuppertal 1999  \cite{Wuppertal},
{\bf 25:}     Seattle 2000 \cite{Seattle},
{\bf 26:}     Sevres 2001  \cite{Sevres01}, 
{\bf 27:}     Lake Brasimone 2001 \cite{lake Brasimone}. 
  Data compilation  adapted from \cite{MMGdata}. }   
\label{fig:GData}}\end{figure}

From these considerations it
follows that the measurement of the value of $G$ will be difficult as the measurement of the  forces
between two of more objects, which is the usual method of measuring $G$, will depend on the geometry
of the spatial positioning of these objects  in a way not previously accounted for because the
Newtonian Inverse Square Law has always been assumed, or in some case  a specified change in the
form of the law has been used.  But in all cases a `law' has been assumed, and this may have been
the flaw in  the analysis of data from such experiments.  This implies that the value of
$G$ from such experiments will show some variability as a systematic effect has  been neglected in
analysing the experimental data, for in none of these experiments is spherical symmetry present.  So
experimental measurements of $G$  should show an unexpected contextuality.  As well the influence of
surrounding matter has also not been properly accounted for. Of course  any effects of turbulence in
the inhomogeneities of the flow has presumably also never even been contemplated.  
The first measurement of $G$ was in 1798 by Cavendish using a torsional balance.  As the precision of
experiments increased over the years and a variety of techniques used  the disparity between the
values of $G$ has actually increased \cite{Gillies}.  Fig.\ref{fig:GData} shows the results from precision 
measurements of
$G$ over the last 60 years. As can be seen  one indication of the contextuality is that measurements of $G$  produce
values that differ by nearly 40 times their individual error estimates. In 1998 CODATA increased the uncertainty in
$G$ from 0.013\% to 0.15\%.  It is predicted that these $G$ anomalies will only be resolved when  the new theory
of gravity is used in analysing the data from these experiments.

  There are  additional gravitational anomalies that are not well-known in physics, presumably because their existence is
incompatible with the Newtonian or the  Hilbert-Einstein gravity theories.
The most significant of these anomalies is the Allais effect \cite{Allais, Allais2}.   In June 1954 Allais
reported that a short Foucault pendulum, known as a paraconical pendulum, exhibited peculiar rates of
precession  at the time of a solar eclipse.  Allais was recording the precession of the pendulum in
Paris. Coincidently during the 30 day observation period a partial solar eclipse occurred at Paris on June 30. 
During the eclipse the precession of the pendulum was seen to be disturbed.  Similar results were obtained during
another solar eclipse on October 29 1959.  There have been other repeats of the Allais experiment with varying
results.  

Another anomaly was reported by Saxl and Allen \cite{Saxl} during the solar eclipse of March 7 1970.  Significant
variations in the period of a torsional pendulum were observed  both during the eclipse and as well in the hours just
preceding and just following the eclipse.  The effects seem  to suggest that an ``apparent wavelike structure has been 
observed over the
course of many years at our Harvard laboratory'', where the wavelike structure is present and reproducible even in the
absence of an eclipse. 

Again Zhou and Huang \cite{Zhou}  report various time anomalies occuring during the solar eclipses of September 23
1987,  March 18 1988 and  July 22 1990 observed using atomic clocks.

All these anomalies and others not discussed here would suggest that  gravity has aspects to it that are not within
the prevailing theories, but that the in-flow theory discussed above might well provide an explanation, and indeed
these anomalies may well provide further phenomena that could be used to test the new theory.  The effects  associated
with the solar eclipses could presumably follow from the alignment of the sun, moon and the earth causing
enhanced turbulence.  The Saxl and Allen experiment of course suggests, like the other experiments analysed
in \cite{AMGE}, that the turbulence is always present. To explore these anomalies detailed numerical studies of
(\ref{eqn:f3vorticitya})-(\ref{eqn:f3vorticityb}) are underway with particular emphasis on the effect of the position of the moon.

\vspace{3mm}

\section{ Galactic In-flow  and the CMB Frame\label{section:galacticinflow}}
  
Absolute motion (AM) of the Solar system has been observed in the direction
$(\alpha=17.5^h,\delta=65^0)$, up to an overall sign to be sorted out,  with a speed of
$417 \pm 40$ km/s. This is the velocity after  removing the contribution of the earth's
orbital speed and the sun in-flow effect \cite{NovaBook,AMGE}. It is significant that this velocity is different
to that associated with the Cosmic Microwave Background 
(CMB) relative to which the Solar system
has a speed of $369$ km/s in the direction
 $(\alpha=11.20^h,\delta=-7.22^0)$, see \cite{CMB}. 
This CMB velocity is obtained by finding the preferred frame in which this thermalised
$3^0$K radiation is isotropic, that is by removing the dipole component.  
The CMB velocity is a measure of the motion
of the Solar system relative to the universe as a whole, or atleast a shell of the universe
some 14Gyrs away, and indeed the near uniformity of that radiation in all directions
demonstrates that we may  meaningfully refer to the spatial structure of the
universe.  The concept here is that at the time of decoupling of this radiation from
matter that matter was on the whole, apart from small observable fluctuations, at
rest with respect to the quantum-foam system that is space. So the CMB velocity is the
motion of the Solar system  with respect to space {\it universally},  but not
necessarily with respect to the   {\it local} space.  Contributions to this  global CMB velocity 
 arise from the orbital motion of the earth in the solar system (this contribution is apparent in the CMB
observational data and is actually removed in the analysis), the orbital motion of the Solar system within the Milky
Way galaxy, giving  a speed of some 230 km/s giving together with  local motion of the Solar
system  in the Milky Way,   a net speed of some 250 km/s,   and contributions from the motion
of the Milky Way within the local cluster, and so on to perhaps larger clusters.

On the other hand the AM velocity is a vector sum of this {\it global} 
velocity and the net velocity associated with the {\it local} gravitational in-flows into
the Milky Way and into the local cluster.  This is because the observation of the CMB velocity does not pick up the
local gravitational in-flows.  Only gravitational lensing could affect that result, and that is an extremely small
effect within the Milky Way.   If the CMB velocity had been identical to the AM velocity then the in-flow 
interpretation of gravity would have been proven wrong. We therefore have three pieces of experimental evidence for
this interpretation (i) the refractive index anomaly discussed previously in connection with the Miller data, (ii)
the turbulence seen in all detections of absolute motion, and now (iii) that the AM velocity is different in both
magnitude and direction from that of the  CMB  velocity. 

That the AM and CMB velocities are different contributes to the explanation offered herein for the resolution  of
the `dark matter' problem. Rather than the galactic velocity anomalies being caused by  undiscovered `dark
matter' we see that the in-flow into non spherical galaxies, such as the spiral Milky Way, will be
non-Newtonian.   As well it will be interesting to determine, at least theoretically, the scale of turbulence
expected in galactic systems, particularly as the magnitude of the turbulence seen in the AM velocity is somewhat
larger than might be expected from the sun in-flow alone. Any theory for the turbulence effect will certainly be
checkable within the Solar system as the time scale of this is suitable for detailed observation.

\section{ Gravitational Waves  \label{section:igravitationalwaves}}

The velocity  flow-field equation is
expected to have solutions possessing turbulence, that is,  fluctuations in both the
magnitude and direction of the gravitational in-flow component of the velocity flow-field.   
Indeed all the gas-mode Michelson interferometer experiments and coaxial cable experiments showed evidence of such
turbulence. The first clear evidence was from the Miller experiment, as shown in the analysis in \cite{NovaBook,AMGE}.   
Miller offered no explanation for these fluctuations  but in his analysis of that data he did running time
averages.   Miller may have in fact have simply
interpreted these fluctuations as purely instrumental effects.  While some of these fluctuations may be partially
caused by weather related temperature and pressure  variations, the bulk of the fluctuations appear to be larger
than expected from that cause alone.   Even the original Michelson-Morley data in
shows variations in  the velocity field and supports this
interpretation.    However it is significant that the non-interferometer DeWitte \cite{AMGE} data also shows
evidence of turbulence in both the magnitude and  direction of the velocity flow field.  Just as the DeWitte data agrees
with the Miller data for speeds and directions the magnitude fluctuations are very similar in absolute magnitude.

It therefore  becomes clear that there is
strong evidence for these fluctuations being evidence of  physical turbulence in the flow
field.  The magnitude of this turbulence appears to be somewhat larger than that which would be
caused by the in-flow of quantum foam towards the sun, and indeed following on from
Sect.\ref{section:galacticinflow}  some of this turbulence may be associated with galactic
in-flow into the Milky Way.  This in-flow turbulence is a form of gravitational wave and the
ability of gas-mode Michelson interferometers to detect absolute motion means that experimental
evidence of such a wave phenomena has been available for a considerable period of time.

\section{ Conclusions\label{section:conclusions}}

Previous analysis \cite{NovaBook,AMGE} of extensive data from both interferometric and non-interferometric experiments has
produced distinctive evidence  for the existence of a quantum-foam substratum to space. 
Effects of motion through this substratum as well as flows related to gravity are evident in this experimental data.  The
evidence suggests that in fact the special relativity effects, which are well established by experiment, are being caused by
absolute motion of systems through this quantum foam  that is space. Process Physics in conjunction with this data leads to a new
theory of gravity  which is  shown to be mathematically consistent with the Newtonian and General Relativity theories in
those cases where these theories have been thoroughly tested. The new theory of gravity has a fundamentally different interpretation
and ontology.   However the new theory of gravity implies that the Newtonian theory of gravity is only strictly applicable to
cases of high spherical symmetry, and that this limitation of the Newtonian theory was inherited by General Relativity in its
formulation by Hilbert and Einstein. The failure of these theories in cases of highly non-spherical systems, such as spiral
galaxies, has resulted in the spurious introduction of concepts like `dark matter'. However in the case of the Global
Positioning System the earth-satellite system  has high spherical symmetry  and in this case the new theory and
General Relativity are mathematically equivalent, and so the obvious success of General Relativity in modelling
the quite complex operations of the GPS is also equally applicable to the new gravity theory.   The failure of
General Relativity in cases of high non-spherical symmetry implies that the explanation of the operation offered
by General Relativity  for the GPS was essentially accidental, and certainly involved an incorrect interpretation
and ontology.

\section{ References\label{section:references}}

\end{document}